\documentclass[11pt,aps,showpacs,superscriptaddress,groupedaddress]{revtex4}
\usepackage{graphicx}
\usepackage{bbm}
\usepackage{mathtools}

\rightline{Date: \today}

\begin{document}


\title{Observing Geometric Torsion}

\author{Stefano Lucat\footnote{Electronic address: {\tt s.lucat@uu.nl}}}

\author{Tomislav Prokopec\footnote{Electronic address: {\tt t.prokopec@uu.nl}}}

\address{\it Institute for Theoretical Physics, Spinoza Institute and EMME$\Phi$, Utrecht University,
Postbus 80.195, 3508 TD Utrecht, The Netherlands}

\begin{abstract}

Dynamical (propagating) torsion can be observed by using conventional gravitational wave detectors such as LIGO, Virgo, LISA
and bar detectors.
We discuss specific signatures of different types of torsion, in particular those of vector and  
mixed symmetric torsion (skew symmetric torsion cannot be detected in this way). 
These signatures are specific to torsion and therefore they can be unambiguously distinguished from those of gravitational waves.

\end{abstract}

\maketitle



\setcounter{page}{1}



\section{Introduction}
\label{Introduction}

 Cartan-Einstein (CE) theory is a very old topic (for reviews that are still actual see~\cite{Hehl:1976kj,Shapiro:2001rz}), 
and yet as far as we are aware of there has been no proposal to detect (dynamical) torsion
{\it via} instruments such as gravitational wave detectors. In this letter we argue that conventional gravitational wave detectors can be used to 
detect propagating (dynamical) torsion. The probable reasons why this has not been proposed before are (a) skew symmetric torsion
(which is typically what one means by torsion in Cartan-Einstein theory) does not imprint any signal on gravitational wave detectors
and (b) torsion is not dynamical in Cartan-Einstein theory. 

If Weyl symmetry is realized at the classical level then torsion trace vector couples to scalar fields, 
implying that scalars source torsion trace~\cite{Lucat:2016eze}, modifying thus the original Cartan-Einstein theory. 
Furthermore, in the original CE theory torsion figures as a constraint (non-dynamical) field, which exists locally where the source
is, but does not propagate. However, when matter is quantized, one can show that (when one integrates out matter fields) already 
at the one-loop level, both torsion trace and skew symmetric 
torsion become dynamical~\cite{Shapiro:2001rz,LucatProkopec:2017}, 
such that they can propagate through space in form of torsion waves and carry energy and information throughout 
our universe just as gravitational waves do. We do not know whether this is realized in nature, but if true the possibilities are exciting enough 
to deserve a closer attention of both theorists and observers. 

Torsion has been used in literature for various purposes. For example, torsion was proposed as a way of avoiding 
the Big Bang singularity \cite{Poplawski:2011jz,Poplawski:2012ab,Lucat:2015rla}, 
to drive inflation~\cite{Boehmer:2005sw,Wang:2008qp,Poplawski:2010kb} and create perturbations~\cite{Wu:2016dkt} 
or primordial magnetic fields~\cite{GarciadeAndrade:2012zz}, 
to generate dark matter~\cite{Belyaev:2016icc} or dark energy~\cite{Wanas:2010xt,Poplawski:2011wj}.
Apart from being detectable by gravitational wave detectors, torsion might be also detectable at the LHC~\cite{Belyaev:2007fn}.


\section{Detection}
\label{Detection}

Just like in general relativity, where gravitational waves induce a change in distance between two test bodies described by the Jacobi equation,
in a more general geometric theory that contains geometric torsion the suitably generalized Jacobi equation~\cite{Lucat:2016eze}
 governs the distance between two test bodies.
In Ref.~\cite{Lucat:2016eze} we have shown that the Jacobi equation reads,
\vskip -0.6cm
\begin{equation}
\nabla_{\dot\gamma}\nabla_{\dot\gamma} J + 2 \nabla_{\dot\gamma} T(\dot\gamma,J) = R(\dot\gamma,J)\dot\gamma
\,,
\label{Jacobi equation}
\end{equation}
\vskip -0.2cm\noindent
where $T(\cdot,\cdot)$ and $R(\cdot,\cdot)$ denote the torsion and curvature tensors, respectively, and
$\nabla_{\dot\gamma}$ is the covariant derivative in the direction of the tangent vector $\dot\gamma$. Usually Jacobi vector fields $J$
are taken to be orthogonal to $\dot\gamma$ ($g(J,\dot\gamma)=0$, where $g(\cdot,\cdot)$ is the metric tensor), but that in fact is not necessary.
In Appendix~A we present details on how torsion enters Eq.~(\ref{Jacobi equation}). Here it is important to keep in mind that
- according to the Young classification of tensors - the torsion tensor $T$ can be decomposed into three distinct tensors,
%
\begin{enumerate}
\vskip -0.7cm
\item torsion trace vector ${\cal T}$;
\item skew symmetric torsion  $\Sigma$;
\item mixed torsion tensor $Q$.
\end{enumerate}
The precise relation is given by Eq.~(\ref{A:torsion tensor: decomposition}) in Appendix~A. 
%
%
When this decomposition and Eqs.~(\ref{A:connection}--\ref{contorsion tensor}) are used in~(\ref{Jacobi equation}), 
one gets how different torsion components contribute to the Jacobi field acceleration. In particular, 
for torsion trace contributes as~(\ref{A:Jacobi equation:torsion trace}), 
skew symmetric torsion as~(\ref{A:Jacobi equation:skew symmetric})
and finally mixed torsion as~(\ref{A:Jacobi equation:mixed}). In what follows we analyze how different
torsion components influence the distance $J$ between neighboring geodesics. For completeness we first present 
the well known result for gravitational waves.

In current earthly measurements (and planned measurements in space) any perturbations of spacetime can be viewed as a small perturbation 
away from Minkowski metric,
$\eta_{\mu\nu}={\rm diag}(-1,1,1,1)$. Throughout this work we assume that both gravitational metric perturbations 
$h_{\mu\nu}(x) = g_{\mu\nu}(x)-\eta_{\mu\nu}$ and torsion perturbations $T^\alpha_{\;\;\mu\nu}(x)$ are small such that 
we can linearize in $h_{\mu\nu}$ and in $T^\alpha_{\;\;\mu\nu}$. In this linear approximation, 
to the required accuracy one can set, $\dot\gamma^\mu=(1,0,0,0)^T$ and $\dot\gamma_\mu=(-1,0,0,0)$.

{\bf Gravitational waves.} To detect gravitational waves it is convenient to work in traceless, transverse (TT) gauge, in which 
$h_{\mu 0} =0$, $\delta_{ij} h_{ij}=0$ and $\partial_i h_{ij}=0$. This is also known as the physical gauge because in this gauge $h_{ij}$ is 
(gauge) invariant to linear coordinate shifts $\xi^\mu$, {\it i.e.} ${\cal L}_\xi h_{ij}=0$, such that in this gauge $h_{ij}$ is a physical (measurable)
quantity. From Eqs.~(\ref{Jacobi equation}) and~(\ref{curvature tensor})
we know that only $\overset{\!\circ}{R^i}_{00j}$ components of the Riemann tensor
contribute. Next, in TT gauge $\overset{\!\circ}{R^i}_{00j}=(1/2)\ddot h_{ij}(t,\vec x)$ and 
$ \dot\gamma^\mu\dot\gamma^\nu\overset{\circ}\nabla_\mu\overset{\circ}\nabla_\nu  J^i $ 
can be approximated by $d^2J^i/dt^2$ 
(because the Levi-Civita connection contributes at second order), such that we have, 
\vskip -0.7cm
\begin{equation}
 \frac{d^2 J^i}{dt^2} = \frac12  \ddot h_{ij}(t,\vec x)J^j
\,.
\label{Jacobi equation: gravitational waves}
\end{equation}
\vskip -0.1cm\noindent
Gravitational waves are built from spin two massless gravitons, which come in two polarizations, known as the plus ($+$) and cross ($\times$)
polarization. For example, if the axes are chosen such that a plane gravitational wave is moving in the $z$-direction, 
then $h_{zz} = 0 $ and,
\begin{enumerate}
\item[A)] Plus  polarization: $h_{xx}=-h_{yy}=h_+\cos(\omega t -  kz)$;
\item[B)] Cross polarization: $h_{xy}=h_{yx}=h_\times\cos(\omega t -  kz)$;
\end{enumerate}
\vskip -0.15cm\noindent
where $\omega = ck$ ($k=\|\vec k\,\|$) is the frequency of the wave  
and $h_+$ and $h_\times$ denote the amplitude of the $+$  and $\times -$polarized wave, respectively. 
By making the {\it Ansatz}, $J^i(t,\vec x) = J^i_{(0)}+\Delta J^i_{(0)}\cos(\omega t - k z)$, one can easily show that 
to leading order in $h_+$ ($h_\times$) equation~(\ref{Jacobi equation: gravitational waves}) is solved by:
\begin{enumerate}
\item[A)] Plus polarization: $J^x(t,z) = J^x_{(0)}\big[1 + (h_+/2)\cos(\omega t -  kz)\big]$, 
$J^y(t,z) = J^y_{(0)}\big[1-(h_+/2)\times\cos(\omega t -  kz)\big]$;
\item[B)] Cross polarization:  $J^x(t,z) = J^x_{(0)} + (h_\times/2) J^y_{(0)}\cos(\omega t -  kz)\big]$, 
                                               $J^y(t,z) = J^y_{(0)} + (h_\times/2)\times J^x_{(0)}\cos(\omega t -  kz)\big]$.
\end{enumerate}
\vskip -0.15cm\noindent
These solutions show that the response displacements $\Delta J^i_{(0)}$ are in phase with the original wave and 
that A) for the plus polarization the relative displacement:  $\Delta L/L=\Delta J^x_{(0)}/J^x_{(0)}=\Delta J^y_{(0)}/J^y_{(0)}=h_+/2$
 is given by 
one half of the wave amplitude, while for B)  for the cross polarization the relative displacement:  
$\Delta L/L=\Delta J^x_{(0)}/J^y_{(0)}=\Delta J^y_{(0)}/J^x_{(0)}=h_+/2$ is also given by 
one half of the wave amplitude but with the axes $x$ and $y$ switched (explaining the name cross polarization).

{\bf Torsion trace.} The contributions to the Jacobi field acceleration from the torsion trace vector is given 
by~(\ref{A:Jacobi equation:torsion trace}). Analogous to the contributions of Christoffel connection, the terms in the second line 
of~(\ref{A:Jacobi equation:torsion trace}) contribute at second order and thus can be neglected 
(this is because $\dot J^\alpha$ is already of the first order in torsion field), such that we have, 
\vskip -0.5cm
\begin{equation}
 \ddot J^0 = 0
\,,\qquad 
 \ddot J^i = J^0 \dot {\cal T}^i + J^j \partial_j {\cal T}^i
\,.
\label{Jacobi equation:torsion trace}
\end{equation}
\vskip -0.2cm\noindent
Now from $g(J,\dot\gamma) = \dot\gamma\cdot J = 0$ and $\dot\gamma^\mu = \delta^\mu_{\;\;0}$ it follows that 
(to this order) 
$J^0=0$
(which is consistent with the first equation in~(\ref{Jacobi equation:torsion trace}))  and the second equation 
in~(\ref{Jacobi equation:torsion trace}) simplifies to,
\begin{equation}
 \ddot J^i = J^j \partial_j {\cal T}^i 
\,.
\label{Jacobi equation:torsion trace:2}
\end{equation}
To facilitate comparison with gravitational waves, we assume that ${\cal T}^i$ can be written as 
a plane wave moving in the $z$-direction, 
\vskip -0.7cm
\begin{equation}
 {\cal T}^i = {\cal T}^i_{(0)}\cos(\omega t - kz)
\,.
\label{torsion trace plane wave}
\end{equation}
\vskip -0.1cm
It is reasonable to assume that torsion wave constitutes a wave of an almost massless field, in which case $\omega \approx ck$
and the waves are to good approximation transverse, meaning that  
${\cal T}^i=({\cal T}^x,{\cal T}^y,{\cal T}^z)$ with $|{\cal T}^z|\ll |{\cal T}^{x,y}|$  in
Eq.~(\ref{torsion trace plane wave}).
\footnote{From the analysis in Appendix~B it follows that one will typically observe ${\cal T}_\mu^T$ which obeys a massive 
vector field (Proca) equation~(\ref{equation of motion torsion trace}), from which one can naturally impose a Lorentz condition 
(obeyed by any massive vector field), 
$\partial_\mu {\cal T}^\mu=0$, or ${\cal T}_0 = - (ck/\omega ) {\cal T}_z$, where for simplicity we have dropped the superscript $T$.
Therefore, even though Eq.~(\ref{Jacobi equation:torsion trace:2}) tells us that we cannot directly observe ${\cal T}_0$, we can imply its value
from the Lorentz condition.
}
Inserting the {\it Ansatz},
\vskip -0.3cm\noindent
\begin{equation}
 J^i(t,z) = J^i_{(0)} + \Delta J^i_{(0)} \sin(\omega t - k z)
\,,
\label{torsion trace plane wave: response}
\end{equation}
\vskip -0.1cm\noindent 
into the Jacobi equation~(\ref{Jacobi equation:torsion trace:2}) and making use of~(\ref{torsion trace plane wave}) results in,  
\vskip -0.3cm\noindent
\begin{equation}
 \Delta J^{i}_{(0)} =-\frac{c^2k}{\omega^2}{\cal T}^{i}_{(0)}J^z_{(0)}
   \approx -\frac{c}{\omega}{\cal T}^{i}_{(0)}J^z_{(0)}
\,.
\label{torsion trace plane wave: response}
\end{equation}
\vskip -0.1cm\noindent
Let us now pause to discuss this result. Notice first that the phase of 
the response~(\ref{torsion trace plane wave: response}) 
is shifted by $\pi/2$ with respect to the phase of the original wave (this is to be contrasted with no phase shift in the case of 
gravitational waves). This phase shift may be difficult to observe, especially because there may be other sources of phase shift (such as 
dispersivity of the medium through which the waves propagate and the massive nature of the torsion trace). 
However, if we have some confidence that gravitational wave and torsion wave come from the same source 
(which can be established by having a directional information) and that dispersive effects of the propagating medium are negligible,
then this phase shift might be observable. 
A second difference is geometric: while gravitational waves induce a response in the relative length along the same transverse direction 
(for plus polarization) or along the opposite, but still transverse, direction (for cross polarization), torsion trace induces a 
response along the transverse direction which is proportional to the longitudinal direction of the instrument. 
Finally third (and probably the most important) difference between the gravitational waves and torsion trace vector signature 
is in that the relative displacement~(\ref{torsion trace plane wave: response}) is inversely proportional to 
the frequency/wave vector (while no frequency dependence is present in the gravitational wave response). This difference may be crucial when 
distinguishing torsion wave signatures from those of gravitational waves.

{\bf Skew symmetric torsion.}
From  Eq.~(\ref{A:Jacobi equation:skew symmetric}) it follows that,
%
$
 \ddot J^i =  0\,,
$
%
which 
immediately implies that skew symmetric torsion cannot be observed by gravitational wave detectors
\footnote{This conclusion could have been reached by a careful look at the derivation of Jacobi equation~(\ref{Jacobi equation})
in Ref~\cite{Lucat:2016eze}, where it is derived by making use of the geodesic equation, $\nabla_{\dot\gamma} \dot\gamma=0$ and 
of equation, ${\cal L}_{\dot \gamma}J=[J,\dot\gamma]=0$. None of these equations contains any dependence on skew symmetric torsion.
}.

{\bf Torsion with mixed symmetry.}  From Eq.~(\ref{A:Jacobi equation:mixed}) we see that, 
\vskip -0.7cm
\begin{equation}
 \ddot J^i = -2\dot Q^i_{\;\;0j} J^j 
\,,
\label{Jacobi equation:mixed torsion}
\end{equation}
where we have assumed that $J^0=0$. Since we do not know any mechanism by which dynamical $Q$ can be generated,
\footnote{Namely, neither matter (scalars, fermions, vectors) nor gravity couples bilinearly  to $Q$  in the Cartan-Einstein theory, implying that 
no kinetic term for $Q$ can be generated at the one-loop order by integrating out matter or gravitational fields.
This is to be contrasted with torsion trace vector and skew symmetric torsion, whose kinetic terms are generated at one-loop order when 
scalar and fermionic matter is integrated out, respectively, thereby making them 
dynamical~\cite{Shapiro:2001rz,LucatProkopec:2017}.}
 we cannot be sure 
how the wave equation for $Q$ looks like. 
Therefore the analysis presented in what follows represents an educated guess. 
It is reasonable to assume that -- just as any massless waves -- the $Q$ waves are transverse, motivating the following 
{\it Ansatz},
\vskip -0.6cm
\begin{equation}
Q^i_{\;\;0j} = Q_{ij(0)}\cos(\omega t-kz)
\,, \quad \delta_{ij}Q_{ij(0)} = 0 
\,,\quad \omega = ck
\label{ansatz for Q wave}
\end{equation}
where $Q_{ij(0)}$ is a real traceless $3\times 3$ matrix. It is convenient to decompose 
 $Q_{ij(0)}$) into its symmetric and traceless ($S^Q_{ij(0)}$) and antisymmetric  ($A^Q_{ij(0)}$) parts, 
 $Q_{ij(0)}=S^Q_{ij(0)}+A^Q_{ij(0)}$ .
Inserting this into~(\ref{Jacobi equation:mixed torsion})
and assuming  the form for $J^i$ as in~(\ref{torsion trace plane wave: response})
 results in,
\vskip -0.5cm
\begin{eqnarray}
 \Delta J^i_{(0)} \!&=&\! -\frac{2c}{\omega}\Big[S^Q_{ij(0)}+A^Q_{ij(0)}\Big] J^j_{(0)}
\,.\quad
\label{Jacobi equation:mixed torsion:2}
\end{eqnarray}
We thus see that the response to a passing $Q$ wave is much richer than that of a gravitational wave.
To get a better feeling on what~(\ref{Jacobi equation:mixed torsion:2}) tells us, let us assume that 
$Q$ is an almost massless field, in which case we have approximately, $Q_{zi(0)}=0= Q_{iz(0)}$
and the only non-vanishing components 
in~(\ref{Jacobi equation:mixed torsion:2}) are $S^Q_{xx(0)}=-S^Q_{yy(0)}$, $S^Q_{xy(0)}$ and $A^Q_{xy(0)}$,
implying that $S^Q$ resembles gravitational wave, and $A^Q$ has no gravitational analogue.
However, there are important differences:
(A) a phase shift of $\pi/2$ characterizes the $Q$ wave response~(\ref{Jacobi equation:mixed torsion:2}) and 
(B) the response to a $Q$ wave is inversely proportional to frequency.

\section{Summary and discussion}
\label{Summary and discussion}

In this letter we show that dynamical (propagating) torsion can be observed by conventional gravitational wave detectors. 
More precisely, torsion trace and torsion of mixed symmetry can be observed, while skew symmetric torsion cannot.
Roughly speaking, the relative amplitude change due to passage of a torsion wave is,  $\Delta L/L\sim T/k\sim cT/\omega$, where 
$T$ is the torsion wave amplitude and $\omega$ and $k$ denote its frequency and wave number, respectively
(more precise results are given in Eqs.~(\ref{torsion trace plane wave: response}) and~(\ref{Jacobi equation:mixed torsion:2})). 
This is to be contrasted with the response to 
gravitational waves of amplitude $h$, which is of the form, $\Delta L/L\sim h$. This difference
suggests that torsion waves of larger wave length (smaller frequency) will be easier observed, and hence 
instruments in space such as LISA, pulsar timing arrays and measurements of cosmic microwave background are generally 
more sensitive to torsion waves. Furthermore, torsion waves induce a response that is delayed by a quarter period, while 
no such phase shift is present for gravitational waves. Finally, torsion polarization differs from that of gravitational waves,
see Eqs.~(\ref{torsion trace plane wave: response}) and~(\ref{Jacobi equation:mixed torsion:2}).

 In this letter we do not discuss production of torsion waves. Since we assume torsion to be dynamical, 
the usual suspects -- inflation, preheating, phase transitions and violent astrophysical events (such as black hole collisions,
collisions of other compact stellar objects, supernovae, etc.) -- can be invoked to be responsible for production. 
Because the response is more sensitive at low frequencies,
one does not need to produce in inflation a (nearly) scale invariant spectrum to make it observable, implying that a  
blue spectrum of torsion waves from inflation could be also observable.

{\bf Acknowledgements.} 
We acknowledge financial support from
an NWO-Graduate Program grant
and by the D-ITP consortium, a program of the Netherlands
Organization for Scientific Research (NWO) that is funded by the Dutch Ministry of Education,
Culture and Science (OCW).

\section*{Appendix A}

In this appendix we present some details on how to analyze the Jacobi equation~(\ref{Jacobi equation}).
%
%
Jacobi field $J$ represents a space-time vector field that can be used to characterize the distance between neighboring 
geodesics in a congruence of geodesics and it is therefore useful in determining how gravitational wave detectors respond 
to a passing (gravitational or torsion) wave.

 In a space-time with geometric torsion, metric compatibility condition, $\nabla_\mu g_{\alpha\beta}=0$ implies that 
 the connection associated with the covariant derivative $\nabla$ can be written as,
\begin{equation}
 \Gamma^\alpha_{\;\;\mu\nu} = \overset{\!\!\!\circ}{\Gamma^\alpha}_{\mu\nu}+K^\alpha_{\;\;\mu\nu}
 \label{A:connection}
 \end{equation}
where $\overset{\!\!\!\circ}{\Gamma^\alpha}_{\;\mu\nu}$ is the Christoffel connection (Levi-Civita symbol) 
and $K^\alpha_{\;\mu\nu}$ denotes the contorsion tensor defined as,
\begin{equation}
 K^\alpha_{\;\;\mu\nu} = T^\alpha_{\;\;\mu\nu}+T_{\mu\nu}^{\;\;\;\;\alpha}+T_{\nu\mu}^{\;\;\;\;\alpha}
\label{contorsion tensor}
\end{equation}
where torsion tensor $T^\alpha_{\;\;\mu\nu}$ is defined as the antisymmetric part of the connection, 
$T^\alpha_{\;\mu\nu}=\Gamma^\alpha_{\;[\mu\nu]}$. 
The curvature tensor $R(\cdot,\cdot)$ in~(\ref{Jacobi equation}) can be conveniently written in terms of the 
Riemann curvature tensor, 
$\overset{\!\!\!\circ}{R^\alpha}_{\mu\rho\nu}=\partial_\rho \overset{\!\!\!\circ}{\Gamma^\alpha}_{\mu\nu} 
              - \partial_\nu\overset{\!\!\!\circ}{\Gamma^\alpha}_{\mu\rho}
  + \overset{\!\!\!\circ}{\Gamma^\alpha}_{\sigma\rho}\overset{\!\!\!\circ}{\Gamma^\sigma}_{\mu\nu}
  -\overset{\!\!\!\circ}{\Gamma^\alpha}_{\sigma\nu}\overset{\!\!\!\circ}{\Gamma^\sigma}_{\mu\rho}$,
and the contorsion tensor $K$ as,
\begin{equation} 
 R^\alpha_{\;\;\mu\rho\nu} = \overset{\!\!\!\circ}{R^\alpha}_{\mu\rho\nu}
  + \overset{\circ}\nabla_\rho K^\alpha_{\;\;\mu\nu} - \overset{\circ}\nabla_\nu K^\alpha_{\;\;\mu\rho}
  + K^\alpha_{\;\;\sigma\rho}K^\sigma_{\;\;\mu\nu}-K^\alpha_{\;\;\sigma\nu}K^\sigma_{\;\;\mu\rho}
\,,
\label{curvature tensor}
\end{equation}
where $\overset{\circ}\nabla$ represents the (general relativistic) covariant derivative 
taken with respect to the Christoffel connection $\overset{\circ}\Gamma$.

A 3-indexed tensor such as the torsion tensor $T$ can be decomposed into its trace 
part ${\cal T}$, a skew symmetric part $\Sigma$ (which is antisymmetric in all indices) 
and a mixed part $Q$ (which is symmetric in the first two indices and antisymmetric in its latter two indices),
\begin{equation}
  T^\alpha_{\;\;\nu\gamma} = \delta^\alpha_{\;\;[\nu}{\cal T}_{\gamma]}
                           +\Sigma^\alpha_{\;\;\nu\gamma} + Q^\alpha_{\;\;\nu\gamma}
\,,
\label{A:torsion tensor: decomposition}
\end{equation}
where $Q^\alpha_{\;\;\alpha\gamma}=0$, $Q_{[\alpha\nu\gamma]}=0$.
From~(\ref{contorsion tensor}) and~(\ref{A:torsion tensor: decomposition}) it follows that 
the contorsion tensor can be decomposed as, 
\begin{equation}
  K_{\alpha\nu\gamma} = 2g_{\gamma[\nu}{\cal T}_{\alpha]}
                           +\Sigma_{\alpha\nu\gamma} + Q_{\alpha\nu\gamma}
\,.
\label{A:contorsion tensor: decomposition}
\end{equation}

When~(\ref{A:torsion tensor: decomposition}) and~(\ref{A:contorsion tensor: decomposition})
are inserted into~(\ref{Jacobi equation}) one obtains contributions from various components 
of the torsion tensor to acceleration of the Jacobi field along $\gamma$. To leading order in torsion these read, 
\begin{eqnarray}
 \bigg(\dot\gamma^\mu\dot\gamma^\nu\overset{\circ}\nabla_\mu\overset{\circ}\nabla_\nu  J^\alpha \bigg)_{\cal T}
 &=& \Big(\overset\circ\nabla_J{\cal T}^\alpha\Big)
        + \dot\gamma^\alpha\Big(\overset\circ\nabla_J{\cal T}_{\dot\gamma}\Big)
\label{A:Jacobi equation:torsion trace}\\
&&  \hskip -2cm
   -2{\cal T}^\alpha\Big(\frac{\overset\circ{\! D}}{D\tau}J_{\dot\gamma}\Big)
        -\Big(\overset\circ\nabla_{\cal T}J^\alpha\Big)
         +\dot\gamma^\alpha {\cal T}_\rho\Big(\frac{\overset\circ{\! D}}{D\tau}J^\rho\Big)
\,,
\nonumber
\\
 \bigg(\dot\gamma^\mu\dot\gamma^\nu\overset{\circ}\nabla_\mu\overset{\circ}\nabla_\nu  J^\alpha \bigg)_{\Sigma}
&=& 0 
\,,
\label{A:Jacobi equation:skew symmetric}
\\
\bigg(\dot\gamma^\mu\dot\gamma^\nu\overset{\circ}\nabla_\mu\overset{\circ}\nabla_\nu  J^\alpha\bigg)_{Q}
&=&-2\dot\gamma^\mu\Big(\frac{\overset\circ{\! D}}{D\tau}Q^\alpha_{\;\;\mu\rho}\Big)J^\rho
\,,
\label{A:Jacobi equation:mixed}
\end{eqnarray}
where 
$J_{\dot\gamma}=\dot\gamma^\mu J_\mu$, ${\cal T}_{\dot\gamma}=\dot\gamma^\mu {\cal T}_\mu$,
$\overset\circ\nabla_J=J^\mu\overset\circ\nabla_\mu$ 
and $\overset{\circ\!} D/D\tau \equiv \dot\gamma^\mu \overset\circ\nabla_\mu$. 
These results are used in the main text to investigate how one can detect torsion waves induced by 
the torsion trace ${\cal T}$, skew symmetric torsion $\Sigma$ or mixed torsion $Q$.


\section*{Appendix B}


In this appendix we discuss propagation of torsion trace, 
as it is implied by the effective gravitational action obtained upon integrating 
out scalar matter~\cite{LucatProkopec:2017}. 
We shall not discuss here skew symmetric torsion, since it is not detectable by standard gravitational wave instruments. 
The equation of motion for dynamical torsion trace is of the form~\cite{LucatProkopec:2017}, 
\begin{equation} 
 \theta\nabla^\mu {\cal T}_{\mu\nu} + \nabla_\nu\Big(\alpha R + \xi \phi^2\Big)  = 0 
\,,
\label{equation of motion torsion trace}
\end{equation}
where $\theta$, $\alpha$ and $\xi$ are (dimensionless) coupling constants, $R$ is the curvature scalar, 
$\phi$ is the (dilaton) scalar
and  ${\cal T}_{\mu\nu}=\partial_\mu{\cal T}_\nu -\partial_\nu{\cal T}_\mu $ is the torsion trace field strength.
To study propagation in the late universe one can replace covariant derivatives with partial derivatives and the metric tensor with 
Minkowski metric $\eta_{\mu\nu}$. The linear form of~(\ref{equation of motion torsion trace}) then becomes, 
\begin{equation} 
 \partial^2 {\cal T}_{\mu} -\Big(1+\frac{6\alpha}{\theta}\Big)\partial_\mu\big(\partial^\alpha {\cal T}_\alpha\big)
    +\frac{2M_{\rm P}^2}{\theta} {\cal T}_\mu = 0 
\,,\;\; \partial^2=\eta^{\mu\nu}\partial_\mu\partial_\nu,
\label{equation of motion torsion trace:2}
\end{equation}
where we wrote $\xi\phi^2 =M_{\rm P}^2$ is the reduces Planck mass squared and 
we assumed $R\approx 0$. 
From~(\ref{equation of motion torsion trace:2}) we see that ${\cal T}_\mu$ is very massive, unless $\theta$ is very large, which is 
what we ought to assume (in order for ${\cal T}_\mu$ to be able to propagate to large distances). 
That also means that one expects typically, $6|\alpha/\theta|\ll 1$. 
In Ref.~\cite{Shapiro:2001rz} it was argued that ${\cal T}$ must be heavy in order to prevent matter fields to excessively decay into torsion.
In our model however, a large value of $\theta$ (needed to make ${\cal T}$ light) also implies a weak coupling 
($\sim 1/\theta\ll 1$) to matter fields. 
Eq.~(\ref{equation of motion torsion trace:2}) 
resembles that of a massive photon, but with a covariant gauge term added {\it a la\/} St\"{u}ckelberg. 
To analyze~(\ref{equation of motion torsion trace:2}) it is convenient to break  ${\cal T}_\mu$ 
 into transverse and longitudinal components,
%
$
 {\cal T}_\mu = {\cal T}_\mu^T + {\cal T}_\mu^L
\,,{\cal T}_\mu^L =( \partial_\mu/\partial^2) \big(\partial^\alpha {\cal T}_\alpha\big)
\,
$
%
such that~(\ref{equation of motion torsion trace:2}) becomes, 
\begin{eqnarray} 
             \partial^2T_\mu^L-\frac{M_{\rm P}^2}{3\alpha} {\cal T}_\mu^L = 0 
\,,
\,\quad 
           \partial^2 T_\mu^T+\frac{2M_{\rm P}^2}{\theta} {\cal T}_\mu^T = 0 
\,.\quad
\label{equation of motion torsion trace}
\end{eqnarray}
That means that, even though the longitudinal field propagates, it is very heavy and it does not propagate very far. On the other hand, 
if $\theta\gg 1$ the transverse field is light and can propagate far. This is what was assumed 
when we analyzed detection of torsion trace 
in~(\ref{Jacobi equation:torsion trace}--\ref{torsion trace plane wave: response}). 
While the analysis presented in this appendix (based on Ref.~\cite{LucatProkopec:2017}) 
illustrates what kind of equation might govern propagation of torsion trace, 
our analysis of detection is general and thus not limited to the model presented here.



\begin{thebibliography}{99}

\bibitem{Hehl:1976kj}
  F.~W.~Hehl, P.~Von Der Heyde, G.~D.~Kerlick and J.~M.~Nester,
  ``General Relativity with Spin and Torsion: Foundations and Prospects,''
  Rev.\ Mod.\ Phys.\  {\bf 48} (1976) 393.
  doi:10.1103/RevModPhys.48.393

\bibitem{Shapiro:2001rz}
  I.~L.~Shapiro,
  ``Physical aspects of the space-time torsion,''
  Phys.\ Rept.\  {\bf 357} (2002) 113
  doi:10.1016/S0370-1573(01)00030-8
  [hep-th/0103093].


\bibitem{Lucat:2016eze}
  S.~Lucat and T.~Prokopec,
  ``The role of conformal symmetry in gravity and the standard model,''
  Class.\ Quant.\ Grav.\  {\bf 33} (2016) no.24,  245002
  doi:10.1088/0264-9381/33/24/245002
  [arXiv:1606.02677 [hep-th]].

\bibitem{LucatProkopec:2017}
  S.~Lucat and T.~Prokopec, in progress (2017).

\bibitem{Poplawski:2011jz}
  N.~J.~Poplawski,
  ``Nonsingular, big-bounce cosmology from spinor-torsion coupling,''
  Phys.\ Rev.\ D {\bf 85} (2012) 107502
  doi:10.1103/PhysRevD.85.107502
  [arXiv:1111.4595 [gr-qc]].

\bibitem{Poplawski:2012ab}
  N.~J.~Poplawski,
  ``Big bounce from spin and torsion,''
  Gen.\ Rel.\ Grav.\  {\bf 44} (2012) 1007
  doi:10.1007/s10714-011-1323-2
  [arXiv:1105.6127 [astro-ph.CO]].

\bibitem{Lucat:2015rla}
  S.~Lucat and T.~Prokopec,
  ``Cosmological singularities and bounce in Cartan-Einstein theory,''
  arXiv:1512.06074 [gr-qc].

\bibitem{Boehmer:2005sw}
  C.~G.~Boehmer,
  ``On inflation and torsion in cosmology,''
  Acta Phys.\ Polon.\ B {\bf 36} (2005) 2841
  [gr-qc/0506033].

\bibitem{Wang:2008qp}
  C.~H.~Wang and Y.~H.~Wu,
  ``Inflation in R + R**2 Gravity with Torsion,''
  Class.\ Quant.\ Grav.\  {\bf 26} (2009) 045016
  doi:10.1088/0264-9381/26/4/045016
  [arXiv:0807.0069 [gr-qc]].

\bibitem{Poplawski:2010kb}
  N.~J.~Popławski,
  ``Cosmology with torsion: An alternative to cosmic inflation,''
  Phys.\ Lett.\ B {\bf 694} (2010) 181
   Erratum: [Phys.\ Lett.\ B {\bf 701} (2011) 672]
  doi:10.1016/j.physletb.2010.09.056, 10.1016/j.physletb.2011.05.047
  [arXiv:1007.0587 [astro-ph.CO]].

\bibitem{Wu:2016dkt}
  Y.~P.~Wu,
  ``Inflation with teleparallelism: Can torsion generate primordial fluctuations without local Lorentz symmetry?,''
  Phys.\ Lett.\ B {\bf 762} (2016) 157
  doi:10.1016/j.physletb.2016.09.025
  [arXiv:1609.04959 [gr-qc]].


\bibitem{GarciadeAndrade:2012zz}
  L.~C.~Garcia de Andrade,
  ``Primordial magnetic fields and dynamos from parity violated torsion,''
  Phys.\ Lett.\ B {\bf 711} (2012) 143.
  doi:10.1016/j.physletb.2012.03.075


\bibitem{Belyaev:2016icc}
  A.~S.~Belyaev, I.~L.~Shapiro and M.~C.~Thomas,
  ``Torsion as a Dark Matter Candidate from the Higgs Portal,''
  arXiv:1611.03651 [hep-ph].

\bibitem{Wanas:2010xt}
  M.~I.~Wanas,
  ``Dark Energy: Is It of Torsion Origin?,''
  arXiv:1006.2154 [gr-qc].

\bibitem{Poplawski:2011wj}
  N.~J.~Poplawski,
  ``Four-fermion interaction from torsion as dark energy,''
  Gen.\ Rel.\ Grav.\  {\bf 44} (2012) 491
  doi:10.1007/s10714-011-1288-1
  [arXiv:1102.5667 [gr-qc]].

\bibitem{Belyaev:2007fn}
  A.~S.~Belyaev, I.~L.~Shapiro and M.~A.~B.~do Vale,
  ``Torsion phenomenology at the LHC,''
  Phys.\ Rev.\ D {\bf 75} (2007) 034014
  doi:10.1103/PhysRevD.75.034014
  [hep-ph/0701002].

\end{thebibliography}
\end{document}